\input definit.tex
\documentstyle[12pt]{article}
\def\beq{\begin{equation}}
\def\eeq{\end{equation}}
\def\bea{\begin{eqnarray}}
\def\eea{\end{eqnarray}}
\def\bq{\begin{quote}}
\def\eq{\end{quote}}

\def\gappeq{\mathrel{\rlap {\raise.5ex\hbox{$>$}}
{\lower.5ex\hbox{$\sim$}}}}

\def\lappeq{\mathrel{\rlap{\raise.5ex\hbox{$<$}}
{\lower.5ex\hbox{$\sim$}}}}

\begin{document}
\topmargin -0.5cm
\oddsidemargin +0.2cm
\evensidemargin -1.0cm
\pagestyle{empty}
\begin{flushright}
{CERN-TH.7094/93}\\
PM 93/41
\end{flushright}
\vspace*{5mm}
\begin{center}
{{\bf PRECISE DETERMINATION OF
{\boldmath $f_{P_{S}}/f_P$}} \
{\bf AND MEASUREMENT OF THE ``PERTURBATIVE'' POLE MASS  FROM
{\boldmath $f_P$}}  } \\
\vspace*{1cm}
{\bf S. Narison} \\
\vspace{0.3cm}
Theoretical Physics Division, CERN\\
CH - 1211 Geneva 23\\
and\\
Laboratoire de Physique Math\'ematique\\
Universit\'e de Montpellier II\\
Place Eug\`ene Bataillon\\
34095 - Montpellier Cedex 05\\
\vspace*{1.0cm}
{\bf ABSTRACT} \\ \end{center}
\vspace*{2mm}
\noindent
We present a compact analytic two-loop
expression of the $SU(3)_f$ breaking effects
on the ratio of the pseudoscalar
decay constants $R_P \equiv f_{P_S}/f_P$ ($P \equiv D,B $),
from which,
we extract
the precise values : $R_D =
1.15 \pm .04$, $R_B = 1.16 \pm .05 $, where the errors are
mainly due to the uncalculated $\alpha_s^2$-corrections.
We also scan carefully the
$M_c$ (resp. $M_b$) mass dependence of
$f_D$ (resp.
$f_B$)
in view of
precise measurements of
the ``perturbative'' c and b pole quark masses from the future data
of the decay constants $f_P$.

\vspace*{2.5cm}
\noindent
%\rule[.1in]{15.0cm}{.002in}

 \vspace*{1.5cm}

\begin{flushleft}
CERN-TH.7094/93 \\
PM 93/41\\
November 1993
\end{flushleft}
\vfill\eject
%\end{document}
\pagestyle{empty}
%\clearpage\mbox{}\clearpage

\setcounter{page}{1}
\pagestyle{plain}

\section{Introduction}
The determination of the $SU(3)_f$ breaking effects in the light
quark sector is important because of the "anomalously" large value of
the strange quark current mass compared to the up and down ones.
As emphasized, a long-time ago by Gell-Mann,
the fact that $m_s \simeq \Lambda_{QCD}$, can lead to
unexpected results
in the strange quark sector.
Indeed, in this sector, the quadratic mass $m_s^2$ corrections can be
as large as the linear one if one uses the OPE of the two-point
correlator for estimating this effect. The linear term
appears only
in the dimension four operator through the $m_s<\bar ss>$
condensate and is suppressed by one power of $q^2 \simeq 1 $GeV$^2$
compared with the $m^2_s$ terms.
These results have been
explicitly verified,
for instance, in the QCD spectral sum rule (QSSR) derivation
of the well-known Gell-Mann-Okubo mass formulae for the light
mesons \cite{SN1} and in the estimate of the ratio of the light quark
condensates $<\bar ss>/<\bar dd>$ \cite{SN2}, \cite{SN3}. However,
we do not yet have a systematic inclusion of this quadratic term in
the present effective chiral Lagrangian approach
\cite{GLE}--\cite{EDR}.Then,
it is not yet possible to test independently
these effects  obtained from QSSR.
\par For the same previous reasons,
it is also interesting to study the $SU(3)_f$ breaking effects
in the heavy quark sector and, in particular, for the heavy-light
mesons systems which possess both symmetries of light and heavy
quarks. Indeed, it is important to verify if the $SU(3)_f$ breaking
corrections are all of the types $1/M_Q$
and/or if there are
other types of important corrections which do not vanish when $M_Q
\rightarrow \infty $.
Moreover, the experimental progress in
measuring the decay constants of the $D_S$ and of other heavy-light
mesons motivates a clean
theoretical control of the $SU(3)_f$-breaking and of the heavy
quark mass
effects on the theoretical estimates of the decay constants.
\par In this paper, we shall evaluate the $SU(3)_f$-breaking
effects on the
ratios of the decay constants of the heavy-light pseudoscalar mesons
defined as :

\begin{equation}
R_P \equiv \frac{f_{P_S}}{f_P},
\end{equation}
where $P \equiv D,B$. We shall also show in detail the effects of
the c (resp. b) quark masses on the decay constants of the $D$
(resp. $B$)
mesons by working with the most accurate sum rule used in Refs.
\cite{SN4}--\cite{SN9}.
\section{\boldmath {$SU(3)_f$}
breaking on the ratio \boldmath {$R_P$} of the decay constants}
For this purpose, we shall use local duality
and QSSR approaches to the pseudoscalar
two-point function :

\begin{equation}
\Psi_5(q^2)  =  i \int d^4x e^{iqx} <0|T J_5(x)J_5^{\dag}(0)|0>,
\end{equation}
associated to the pseudoscalar current :

\begin{equation}
J_5(x) = (m_q+M_Q):\bar q (i \gamma^5) Q : ,
\end{equation}
where $q \equiv d,s$, $
 Q \equiv c,b $ and $m_q$ (resp $M_Q$) are the light (resp. heavy)
quark masses.
\par For illustrating our discussion, let us start with the bare loop
expression of the spectral function. It reads :
\begin{equation}
\rm {Im} \Psi_5(t)  =  \frac{3}{8 \pi}(m_q+M_Q)^2
\frac{(t-(M_Q-m_q)^2)^2}{t} \sqrt{1- \frac{4m_qM_Q}{t-(m_q-M_Q)^2}}.
\end{equation}
We shall make a $m_q$ expansion of this expression. Then, it
is convenient to introduce the low-energy parameter E or $\omega$
defined as :
\begin{equation}
t  =  (E+M_Q)^2 \ \ \  \rm {or} \ \ \  t  =   M^2_Q + \omega M_Q.
\end{equation}
This way, one gets :
\begin{equation}
\rm {Im} \Psi_5 \left( \omega \right)
 \simeq   \frac{3}{8 \pi} \left( m_q+M_Q \right)^2
\frac{\omega^2}{1+ \frac{\omega}{M_Q}} \left\{
1+2 \frac{m_q}{\omega} \left( 1-
\frac{m_q}{M_Q} \right) - \left( \frac{m_q}{\omega}
\right)^2 \right\} .
\end{equation}
Invoking the quark-hadron duality which equates the decays
of a ``pseudoscalar W'' into a resonance and into a $\bar qQ$ pair,
one obtains the sum rule \cite{ZAL} :
\begin{equation}
\int_{0}^{\omega_c} d \omega \rm {Im} \Psi_5^{res}(\omega)=
\int_{0}^{\omega_c} d \omega \rm {Im} \Psi_5^{\bar qQ}(\omega),
\end{equation}
where $\omega_c$ is the continuum energy. Using a narrow width
approximation for the lowest pseudoscalar state, this sum rule
reproduces the correct mass behaviour of the decay constant\cite{ZAL}
in the chiral $m_q$=0 limit.
Retaining the
quadratic $SU(3)_f$ breaking terms,
one then obtains the
sum rule :
\begin{equation}
R_P \simeq \rho_P
\left\{1+3 \left(\frac{m_s}{\omega_c}\right) \left(1-
 \frac{m_s}{M_Q} \right)
-6\left( \frac{m_s}{\omega_c} \right)^2
- \left( \frac{m_s}{M_Q} \right) \left(1- \frac{m_s}{M_Q}\right)
\right\}.
\end{equation}
The quantity $\rho_P$ is defined as :
\begin{equation}
 \rho_P \equiv \left( \frac{M_P}{M_{P_S}}\right)^2
\left(1+ \frac{m_s}{M_Q} \right)
\end{equation}
The sum rule
explicitly indicates that the $SU(3)_f$ breaking corrections
are of two types: the ones in $m_s/M_Q$ vanish when $M_Q \rightarrow
\infty$, while
the other ones in $m_s/ \omega_c$ contribute even in the infinite mass
limit.
One can deduce the value of $\omega_c$ from that of $E_c$ determined
from a numerical evaluation of the $D$ and $B$ sum rules within
stability criteria \cite{SN5}, \cite{SN6}--\cite{SN8}:
\begin{eqnarray}
E_c^D \simeq (1.08 \pm 0.26) \mbox{GeV}, \nonumber \\
E_c^B \simeq (1.30 \pm 0.10) \mbox{GeV},
\end{eqnarray}
which corresponds to the quark mass values obtained from a global QSSR
fit of different hadronic channels
\cite{SN3}, \cite{SN9} :
\begin{eqnarray}
M_c \equiv M_c(p^2=M_c^2) = (1.47 \pm 0.05) \mbox{GeV}, \nonumber \\
M_b \equiv M_b(p^2=M_b^2) = (4.60 \pm 0.05) \mbox{GeV}.
\end{eqnarray}
These values of $E_c$ give :
\begin{equation}
\omega_c \equiv 2E_c \left(
1+ \frac{E_c}{2M_Q} \right) \simeq (3.1\pm 0.1) \mbox{GeV},
\end{equation}
which is equal to the one obtained  when
one uses the duality constraint for $M_Q \rightarrow \infty$ \cite{ZAL}.
This
result indicates that contrary to E$_c$,
$\omega_c$ is constant for the heavy
quarks. So, it is a convenient parameter expansion in the $SU(3)_f$
breaking
analysis. One can further improve the previous sum rule in (8)
by working with the exponential sum rule :
\begin{equation}
L=\int_{0}^{\omega_c} d \omega e^{-\omega \tau}\rm {Im} \Psi_5^{res}
(\omega),
\end{equation}
where $\tau$ is the Laplace (Borel) non-relativistic
sum rule variable.
Varying $\tau$
(resp. $\omega_c$) in the range (0.8-1.2)
 GeV$^{-1}$(resp. (2.7-3.3)GeV)),
where the $D$ and $B$ sum rules stabilize, we deduce the
improved constraint :
\begin{equation}
R_P^2= \rho^2_P
\left\{1+2(2.2 \pm.2) \left(\frac{m_s}{\omega_c}\right) \left(1-
 \frac{m_s}{M_Q}\right)
-2(8.2 \pm 1.6)\left( \frac{m_s}{\omega_c} \right)^2 \right\},
\end{equation}
where the numerical integrations contain a slight $M_Q$ dependence
from the expression in (6). The errors in the numerical coefficients
are due the ones of $\tau$ and $\omega_c$ but, as we shall see later
on, the main sources of errors on $R_P$ do not come from these terms.
To that order, one can deduce,
for $\rho_D$=.998 and $\rho_B$=.995 :
\begin{equation}
R_D \simeq  1.07 \ \ , \ \ R_B \simeq 1.08,
\end{equation}
which one can compare with the moment sum rule result from a
$1/M_b$ expansion \cite{SN6}, \cite{SN5} :
\begin{equation}
R_B \simeq \left( \frac{M_{B_S}}{M_B}\right)
 \left\{1+ \frac{3}{2} \frac{m_s}{M_b}
\left(1- \frac{M_b^2}{t_c}\right)
 + \frac{4 \pi^2}{M_b^3}<\bar ss -\bar dd> \right\}.
\end{equation}
The moment sum rule gives to leading order in $\alpha_s$ :
\begin{equation}
R_D \simeq  1.13 \ \ , \ \ R_B \simeq 1.11,
\end{equation}
which agrees quite well with the one from the Laplace sum rule.
One can improve again this Laplace sum rule
by including the
radiative corrections to the $m_s$ and $m_s^2$
terms  \cite{BGR} :
\begin{eqnarray}
m_s  \rightarrow  m_s \left\{1+ \left(\frac{\alpha_s}{\pi}\right)
\left(2
\log \frac{\omega_c}{\nu} +\frac{7}{3}\right) \right\}, \nonumber \\
m_s^2  \rightarrow  m_s^2 \left\{ 1+ \left( \frac{\alpha_s}{\pi}
\right) \left(-4
\log \frac{\omega_c}{\nu} -\frac{11}{3}-\frac{4}{9}\pi^2 \right)
\right\},
\end{eqnarray}
where $\nu$ is the subtraction scale of the $\overline {MS}$ scheme.
We estimate the errors due
to the unknown
$\alpha_s^2$ terms from arguments based on
an algebraic growth of the QCD series, which give a
coefficient of about 5 times the $\alpha_s$ one \cite{BNP}.
This feature is also observed in some other QCD examples
\cite{PIV}.
Choosing the
renormalization point at $\tau^{-1}$ as dictated by the RGE obeyed by
the non-relativistic Laplace sum rule \cite{SN8},
these radiative terms induce large corrections which
modify the $m_s$ (resp. $m_s^2$) coefficients by a factor $ 1.69 \pm
0.52$
(resp. 0.43 $\pm 0.43$), where we have used for 4 and
5 flavours the value of the QCD scale in the $\overline {MS}$ scheme
\cite{PDG} :
\begin{equation}
\Lambda_4 \simeq (260 \pm 54) \mbox{MeV} \ \ , \ \
\Lambda_5 \simeq (175 \pm 41) \mbox{MeV}.
\end{equation}
Introducing the value of the running strange quark mass to three loops
\cite{SN9}, \cite{SN3}:
\begin{equation}
\bar m_s(1 \ {\mbox GeV}) =  (159.5 \pm 8.8) \mbox{MeV},
\end{equation}
one obtains by including the $\alpha_s$ corrections :
\begin{equation}
R_D =  1.15 \pm .04   \ \ , \ \
R_B =  1.16 \pm .05   ,
\end{equation}
where the errors come mainly from the uncalculated $\alpha_s^2$ terms.
These results are similar to the lattice results reviewed in
Ref.\cite{LAT}.
It is important to notice that the non-relativistic
Laplace sum rule indicates
that the ratio of the decay constants $R_P$  remains constant
when $M_Q \rightarrow \infty$ due to the dominance
of the $1/\omega_c$ rather than of the $1/M_Q$ corrections :
\begin{equation}
R_P (M_Q \rightarrow \infty) = 1.16 \pm .05.
\end{equation}
We combine the previous results with the known values \cite{SN3},
\cite{SN5}--\cite{SN9} of the $D$ and $B$ decay constants corresponding
to the quark mass values in (11) :
\begin{equation}
f_D = (1.31 \pm .12)f_\pi \ \ , \ \ f_B = (1.60 \pm .26)f_\pi,
\end{equation}
Then, we deduce :
\begin{equation}
f_{D_S} = (1.51 \pm .15)f_\pi \ \ , \ \ f_{B_S} = (1.86 \pm .31)f_\pi,
\end{equation}
which can be tested soon in the forthcoming measurement of $f_{D_S}$
and later on of $f_{B_S}$.
\section{Measurement of the c and b quark pole masses from
the decay constants \boldmath {$f_D$} and \boldmath{ $f_B$}}
Moreover, it has also been
noticed previously \cite{SN4}--\cite{SN9} that
the values of $f_P$ obtained from QSSR have a sizeable dependence
on the input values of the heavy quark ``perturbative'' pole mass or
equivalently on the meson-quark mass difference. This effect
mainly induces a large uncertainty and, among other things, an
apparent discrepancy between different predictions from the QSSR method
\cite{SN3}, \cite{SN4}--\cite{BGR}, \cite{PC2}--\cite{DPA}.
Though we expect that the real values of these masses should be in the
range
given in (11), we also propose that
the future
data on the decay constants can be used to fix in an accurate and
independent way the values of the
c and b quark ``perturbative'' pole masses $M_c \equiv M_c(p^2=M^2_c)$,
$M_b \equiv M_b(p^2=M^2_b)$.
For this purpose, we scan the values of
$f_D$ (resp. $f_B$) for  $M_c$ (resp. $M_b$)
belonging in the
range
larger than the one allowed in (11). In so doing, we work with the
Laplace relativistic sum rule (LRS)
which, like the non-relativistic moment
sum rule,
gives the most precise value
of the decay constants \cite{SN7}--\cite{SN9}. For the LRS, this
precision is due to the fast convergence
of the non-perturbative QCD series and to the fact that
the coefficient of the radiative
correction to the lowest perturbative graph is smaller
than in the non-relativistic (HQET) Laplace sum rule. Also,
the sum rule
scale $\tau^{-1}$ at which the QCD coupling is evaluated is higher
for the LRS.
The precision of the moments sum rule is due to the fact that
the QCD coupling is evaluated at the b-quark mass as dictated by its
RGE \cite{SN8}. The explicit form of the LRS
to two-loop and including the dimension-six non-perturbative condensates
has been
given in \cite{SN6},\cite{SN3}, while the details of the sum rules
analysis can be found in \cite{SN3},\cite{SN6}--\cite{SN8}.
We show the results on $f_P$ versus $M_Q$ in the tables, where we
have used the values of $\Lambda$ in (19) and the following values
of the QCD input parameters for $SU(n)_f$ \cite{SN3}, \cite{SN9} :
\begin{equation}
< \bar dd> = -[
(188.6 \pm 6.6) \mbox{MeV}]^3(-\log (\tau ^{-1}\Lambda ))^
{2/{-\beta _1}} \ \ , \ \  - \beta _1 = \frac {1}{2}(11- \frac {2n}{3}),
\end{equation}
and :
\begin{equation}
<{\alpha}_s G^2 > = (.06 \pm .02) \mbox{GeV}^4 \ \ , \ \
g<\bar d \sigma^{\mu \nu} \frac{\lambda^a}{2}
G^{\mu \nu}_a d > = (.80 \pm .01) \mbox{GeV}^2 <\bar dd>.
\end{equation}
\medskip
$$ \vbox{\alignement
{& \vrule#& \colleft# \colright
& \vrule#& \colleft# \colright
&\vrule#\cr
\noalign{\hrule}
\filetvide& \omit&& \omit&\cr
&$\displaystyle \hfill {M_c [\mbox{GeV}]} \hfill$&&$\displaystyle \hfill
 f_D/f_\pi}
\hfill$&\cr
\filetvide& \omit&& \omit&\cr
\noalign{\hrule}
\filetvide& \omit&& \omit&\cr
&$\displaystyle \hfill 1.20 \hfill$&&$\displaystyle \hfill 1.51\pm 0.08
\hfill$&\cr
\filetvide& \omit&& \omit&\cr
&$\displaystyle \hfill 1.25 \hfill$&&$\displaystyle \hfill 1.51\pm
0.05 \hfill$&\cr
\filetvide& \omit&& \omit&\cr
&$\displaystyle \hfill 1.30 \hfill$&&$\displaystyle \hfill 1.49\pm
0.05 \hfill$&\cr
\filetvide& \omit&& \omit&\cr
&$\displaystyle \hfill 1.35 \hfill$&&$\displaystyle \hfill 1.43\pm
0.05 \hfill$&\cr
\filetvide& \omit&& \omit&\cr
&$\displaystyle \hfill 1.40 \hfill$&&$\displaystyle \hfill
1.37\pm 0.04 \hfill$&\cr
\filetvide& \omit&& \omit&\cr
&$\displaystyle \hfill 1.45 \hfill$&&$\displaystyle \hfill 1.30\pm
0.03 \hfill$&\cr
\filetvide& \omit&& \omit&\cr
\filetvide& \omit&& \omit&\cr
&$\displaystyle \hfill 1.50 \hfill$&&$\displaystyle \hfill 1.19\pm
0.02 \hfill$&\cr
\filetvide& \omit&& \omit&\cr
\filetvide& \omit&& \omit&\cr
&$\displaystyle \hfill 1.55 \hfill$&&$\displaystyle \hfill 1.08\pm
0.02 \hfill$&\cr
\filetvide& \omit&& \omit&\cr
%&$\displaystyle \hfill
% \underline{        } \hfill$&&$\displaystyle \hfill
%\underline{           } \hfill$&\cr
%$\displaystyle \hfill R_{\tau ,A} \hfill$&&$\displaystyle \hfill 1.638\pm
%0.033 \hfill$&\cr
\filetvide& \omit&& \omit&\cr
\noalign{\hrule}}}
 $$
\par
\centerline{\underbar{Table 1}}
\medskip
\noindent
The error for $\Lambda$ in (19) induces a correction of 1.5 $\%$.
The ones of the non-perturbative condensates
give corrections, at most, of the
order of $0.6\%$ each. The main errors
in the tables are due the range of $f_P$ corresponding to the
beginning of the sum rule variable stability (low value of $f_P$) until
the beginning of the continuum energy
($E_c$) stability (high value of $f_P$). The central value given in the
tables corresponds to the average of these two extremal values. These
errors can be considered as the systematic errors of the method which
are due to the different criteria for extracting the optimal result.
\medskip
$$ \vbox{\alignement
{& \vrule#& \colleft# \colright
& \vrule#& \colleft# \colright
&\vrule#\cr
\noalign{\hrule}
\filetvide& \omit&& \omit&\cr
&$\displaystyle \hfill {M_b [\mbox{GeV}]} \hfill$&&$\displaystyle \hfill
 f_B/f_\pi}
\hfill$&\cr
\filetvide& \omit&& \omit&\cr
\noalign{\hrule}
\filetvide& \omit&& \omit&\cr
&$\displaystyle \hfill 4.40 \hfill$&&$\displaystyle \hfill 1.86\pm 0.17
\hfill$&\cr
\filetvide& \omit&& \omit&\cr
&$\displaystyle \hfill 4.45 \hfill$&&$\displaystyle \hfill 1.81\pm
0.13 \hfill$&\cr
\filetvide& \omit&& \omit&\cr
&$\displaystyle \hfill 4.50 \hfill$&&$\displaystyle \hfill 1.74\pm
0.09 \hfill$&\cr
\filetvide& \omit&& \omit&\cr
&$\displaystyle \hfill 4.55 \hfill$&&$\displaystyle \hfill 1.66\pm
0.09 \hfill$&\cr
\filetvide& \omit&& \omit&\cr
&$\displaystyle \hfill 4.60 \hfill$&&$\displaystyle \hfill
1.54\pm 0.09 \hfill$&\cr
\filetvide& \omit&& \omit&\cr
&$\displaystyle \hfill 4.65 \hfill$&&$\displaystyle \hfill 1.43\pm
0.05 \hfill$&\cr
\filetvide& \omit&& \omit&\cr
\filetvide& \omit&& \omit&\cr
&$\displaystyle \hfill 4.70 \hfill$&&$\displaystyle \hfill 1.29\pm
0.05 \hfill$&\cr
\filetvide& \omit&& \omit&\cr
\filetvide& \omit&& \omit&\cr
&$\displaystyle \hfill 4.75 \hfill$&&$\displaystyle \hfill 1.18\pm
0.05 \hfill$&\cr
\filetvide& \omit&& \omit&\cr
\filetvide& \omit&& \omit&\cr
&$\displaystyle \hfill 4.80 \hfill$&&$\displaystyle \hfill 1.05\pm
0.05 \hfill$&\cr
\filetvide& \omit&& \omit&\cr
%&$\displaystyle \hfill
% \underline{        } \hfill$&&$\displaystyle \hfill
%\underline{           } \hfill$&\cr
%$\displaystyle \hfill R_{\tau ,A} \hfill$&&$\displaystyle \hfill 1.638\pm
%0.033 \hfill$&\cr
\filetvide& \omit&& \omit&\cr
\noalign{\hrule}}}
 $$
\par
\centerline{\underbar{Table 2}}
\medskip
\noindent
One should not misinterpret the $M_Q$- dependence obtained here for
being in contradiction with the one\cite{SN7},\cite{SN8},\cite{BAG},
\cite{MN1},\cite{LAT} :
\begin{equation}
f_P \sqrt{M_P} \simeq \left(f_P \sqrt{M_P}\right)_\infty
\left(1- \frac{ \left(.88 \pm .18 \right)\mbox{GeV}}
{M_Q} +...\right),
\end{equation}
as in the tables, we vary the quark mass but we keep the meson mass
fixed. One should also notice that for a given value of the quark
mass, the central value presented
here is slightly lower by 3.9$\%$ than the average given in \cite{SN8}.
This difference is due to the fact that the result
from the non-relativistic (HQET) Laplace sum rule taken into account
in the average is large as being due to the
huge value of the $\alpha_s$ correction in the HQET sum rule. This
feature implies a bad
convergence of the perturbative QCD series \cite{BGR}
and then,
a large error from the estimated $\alpha^2_s$
correction \cite{SN8} which restores the agreement of the
HQET results with the previous ones. However,
from these reasons, the HQET sum rule is not
quite adequate for an accurate determination of $f_P$.
The present
lattice results :
\begin{equation}
f_D = (1.51 \pm .23)f_\pi \ \ , \ \ f_B = (1.36 \pm .30)f_\pi,
\end{equation}
as quoted in Ref.\cite{LAT}, are consistent with the ones in (23)
and so favour the values of the pole masses given in (11).
\section{Conclusion}
We have derived analytically the strength of the $SU(3)_f$ breaking
on the ratio of the pseudoscalar decay constants using the two-loop
expression of the pseudoscalar two-point correlator. We have found
that the $SU(3)_f$ breaking corrections are mainly due to the
$1/\omega_c$
terms (Eqs.14, 21 and 22),
which remain constant when $M_Q$ goes to infinity. After the acceptation
of this paper for publication, we have also been aware of the work in
\cite{PAV}, where $R_P$ has also been derived using the usual numerical
procedure of the QSSR approach. We have not used that method in our
paper where an analytic derivation of the SU(3) breaking terms has been
given, which shows in a clean and simple way the different origins of
the SU(3) effects. Moreover, the authors also use uncorrectly the
two-loop QCD expressions given in \cite{25}, which {\it is only valid if
one uses the pole masses for $M_Q$ and $m_s$}. The
definition of $m_s$ used by the authors are inconsistent with this
two-loops expression and may explain the small difference between their
and our results. However, the errors given by the authors are very
underestimated. Our results take into account the errors due to the
uncalculated $\alpha_s^2$ corrections which is not discussed in
\cite{PAV}. The effects of this terms can be larger than the ones
due to the high dimension condensates taken into account in \cite{PAV}.
which give only  negligible SU(3) breaking effects.
\par
\noindent
We have also carefully scanned the mass-dependence of the D and B
decay constants around the c and b quark masses (Tables 1 and 2),
which can be compared with
the forthcoming  data on $f_D$ and $f_B$ in order to deduce an
accurate measurement of the ``perturbative'' quark pole masses
free from theoretical prejudices. Indeed,
this future experimental measurement should help for clarifying
the ``apparent conflict'' between the defenders of low
and those of high values of the heavy quark masses and will serve as a
guide on a deeper theoretical understanding of
the relations between the different quark masses issued from various
analysis of data (QSSR, inclusive heavy meson decays, deep inelastic,
...).
\section*{Acknowledgement}
We thank Toni Pich
for discussions and for
reading the manuscript . Conversations with Ikaros Bigi, Matthias
Jamin and Kacper Zalewski have been also useful.

\end{document}